\documentstyle[psfig]{caosp}
\pubyear{1998}
\volume{27}
\firstpage{171}
\begin{document}
\hauthor{A.E G\'omez {\it et al.}}
%%%%%%%%%%%%%%%%%%%%%%%%%%%%%%%%%%%%%%%%%%%%%%%%%%%%%%%%%%%%%%%%%%%%%%%%%%%%%%
\title{Absolute magnitudes and kinematics of CP stars from Hipparcos data}
\author{A.E. G\'omez \inst{1} \and X. Luri \inst{1,} \inst{2}
\and V. Sabas \inst{1} \and S. Grenier \inst{1} \and F. Figueras \inst{2}
\and P. North \inst{3} \and J. Torra \inst{2} \and M.O. Mennessier \inst{4}}
\institute{Observatoire de Paris-Meudon,
           D.A.S.G.A.L., URA 335 du CNRS,
           F92195  Meudon CEDEX, France
           \and Departament d'Astronomia i Meteorologia,
           Universitat de Barcelona,
           Avda. Diagonal 647, E08028 Barcelona, Spain
           \and Institut d'Astronomie de l'Universit\'e de Lausanne,
           CH1290 Chavannes des Bois, Switzerland
           \and Universit\'e Montpellier II, G.R.A.A.L.,
           URA 1368 du CNRS,
           F34095 Montpellier CEDEX 5, France
          }
%%%%%%%%%%%%%%%%%%%%%%%%%%%%%%%%%%%%%%%%%%%%%%%%%%%%%%%%%%%%%%%%%%%%%%%%%%%%%%
%\date{December 31, 1997}
\maketitle
%%%%%%%%%%%%%%%%%%%%%%%%%%%%%%%%%%%%%%%%%%%%%%%%%%%%%%%%%%%%%%%%%%%%%%%%%%%%%%
\begin{abstract}
The position in the HR diagram and the kinematic characteristics of different
kinds of CP stars of the upper main sequence are obtained using the LM method
(Luri et al., 1996). 
Most of the CP stars are main sequence stars occupying the whole width of the 
sequence. From a kinematic point of view, they belong to the 
young disk population (ages $\la$\, 1.5 Gyr). It has also been found that,
on kinematic grounds, the
behaviour of $\lambda$\,Bootis stars is similar to the one 
observed for normal stars of the same spectral range. On the other hand,
roAp and noAp stars show the same kinematic characteristics.
The peculiar velocity distribution function 
has been decomposed into a sum of three dimensional gaussians 
and the presence of Pleiades, Sirius 
and Hyades moving groups has been clearly established.  
Finally, a small number of CP stars are found to be high-velocity 
objects.
\end{abstract}
%%%%%%%%%%%%%%%%%%%%%%%%%%%%%%%%%%%%%%%%%%%%%%%%%%%%%%%%%%%%%%%%%%%%%%%%%%%%%
\section{Introduction}
\label{intr}
The release of Hipparcos data (ESA, 1997) allows to reconsider
the luminosity of CP stars of the upper main sequence and their kinematic
behaviour
on sounder bases. In the present paper the following kinds of CP stars have 
been considered: He-rich, He-weak, HgMn, Si, SrCrEu and the related group of 
$\lambda$\,Bootis stars.

The LM statistical method (Luri et al., 1996) has been applied to the
different samples. This method has the advantage that
all the available astrometric data (whatever the quality of the parallax is) as
well as radial velocity data are used for  each star for the luminosity
calibration. It also provides the kinematic characteristics of the samples. 
The method is able to
identify and separate groups of stars with different luminosity, kinematic or
spatial properties, allowing the treatement of non homogeneous samples.

\section{Material}

For Bp\,-\,Ap stars, in order to minimize misclassifications, the samples 
have been selected using
different sources and intercomparing the spectral classifications between
them. First, all the stars in Renson's Catalogue (Renson, 1991) observed
by Hipparcos were retained. This first list was cross-correlated
with the Catalogue of Stellar Groups of Jaschek \& Egret (1981) and the Michigan
catalogues: Houk \& Cowley (1975), Houk (1978; 1982) and
Houk \& Smith-Moore (1988). After that, the stars with discrepant
spectral classifications were excluded. The group named Si+ contains
intermediate types like SiCr and SiEu.
For Am stars, two samples have been selected from Hauck's Catalogue
(Hauck, 1992): ``normal'' Am and ``mild'' Am stars. A star was considered
``mild'' when the difference
between the spectral types obtained using metallic lines and the K\,-\,line
of Ca\,{\sc ii} was smaller than 5 subtypes. Table \ref{tab:Ngroups} gives 
the number N of stars in the selected samples and the 
spectral type and effective temperature ranges.
Finally, a sample of 41 $\lambda$\,Bootis stars, taken from the Catalogue of
Pauzen et al. (1997), has also been considered.

%%%%%%%%%%%%%%%%%%%%%%%%%%%%%%%%%%%%%%%%%%%%%%%%%%%%%%%%%%%%%%%%%%%%%%%%%%%%
\begin{table}[h]
 \caption{Samples of Bp-Ap and Am stars}
 \label{tab:Ngroups}
 \small
 \begin{center}
 \begin{tabular}{lrlrr}
   Sample   &   N & Sp. range &  $T_{\rm eff}$ range & N$_f$\\
  \hline
   He-rich &  14 &    B2     & 18000\,-\,23000 K & 14\\
   He-weak &  58 & B4\,-\,B8 & 13000\,-\,17000 K & 58\\
   HgMn    &  76 & B8\,-\,A0 & 10000\,-\,14000 K & 44\\
   Si      & 440 & B7\,-\,A2 &  9000\,-\,14000 K & 415\\
   Si+     &  87 & B8\,-\,A2 &  8000\,-\,13000 K & 66\\
   SrCrEu  & 378 & A0\,-\,F0 &  7000\,-\,10000 K & 353\\
   Am ``normal''&852&A0\,-\,F0 &  7000\,-\,10000 K & 781\\
   Am ``mild''& 207& A0\,-\,F0 &  7000\,-\,10000 K & 184\\
 \hline
 \end{tabular}
 \end{center}
\end{table}
%%%%%%%%%%%%%%%%%%%%%%%%%%%%%%%%%%%%%%%%%%%%%%%%%%%%%%%%%%%%%%%%%%%%%%%
For each star, astrometric data (parallax and proper motion components)
as well as photometric data and  their corresponding errors have been taken
from the Hipparcos Catalogue. Radial velocity data come from
different sources: Barbier-Brossat \& Figon (1997), Dufflot et al.
(1995), Grenier et al. (1997), Levato et al. (1996)
or from Coravel (North, private communication). When a star had more than
one radial velocity source, a mean weighted value was adopted. 

Effective temperatures ($T_{\rm eff}$) were evaluated using Geneva photometry
for all the Bp\,-\,Ap groups
with the exception of the He-rich group. In this last case,
the values given in Zboril et al. (1997) were used. For Am stars, effective
temperatures were obtained from Str\H{o}mgren photometry and for
$\lambda$\,Bootis stars the spectroscopic values from
Cayrel de Strobel et al. (1997) were used.

Absolute magnitudes may be affected by interstellar absorption and duplicity
effects.  The effect of the interstellar absorption on the apparent magnitude
has been taken into account using the tridimensional model of Arenou et
al. (1992) which is included in the LM method. The correction for
binary companions has been applied when the difference of magnitude between the
components was known, otherwise the star was rejected.
 The final number of
stars in the selected samples (N$_{f}$) is given in Table \ref{tab:Ngroups}. 
In the sample of $\lambda$\,Bootis stars 39 stars remained.

\section{The method}

The LM method is based on a Maximum-Likelihood algorithm. Given a selected
sample and a model for the luminosity, the velocity and spatial distributions,
the method:\\
- uses all the available information (astrometric, photometric and
spectroscopic);\\
- takes into account the observational censorship of the sample and the 
  observational errors;\\
- is able to treat a mixture of stars coming from different groups and to
separate them;\\
- provides a mean luminosity calibration as well as individual absolute 
  magnitudes and the spatial and kinematic characteristics of the samples.\\
A normal distribution for the absolute magnitude 
$(M_0,\sigma_M)$, a Schwarzchild ellipsoid ($U_0, V_0, W_0, \sigma_U, 
\sigma_V, \sigma_W$) for the velocity distribution and an exponential-disk
for the spatial distribution ($Z_0$ being the scale height in the direction 
perpendicular to the galactic plane) have been adopted. 
U, V and W are the heliocentric velocity components
in the direction of the galactic center, of the
galactic rotation and of the north galactic pole, respectively. The galactic
differential rotation effect on the kinematic data has been taken
into account using the 
Oort-Lindblad model at first order. Moreover, the sample selection in apparent
magnitude has been described by a selection function which is uniform up to a 
certain magnitude $m_c$ (treated as a parameter to be determined) and then 
linearly decreases up to the limiting magnitude of the sample.

\section{Results}
 
The LM method has been applied separately to the different samples and for
all of them, with the exception of the samples of He-rich, He-weak and
$\lambda$\,Bootis stars, secondary groups were found. Table \ref{tab:resul} 
gives for the main groups, which contain 
the largest number of stars, the dispersion ($\sigma_M$) of the
intrinsic visual absolute magnitude, expressed in 
magnitudes, the velocity dispersions
($\sigma_U, \sigma_V$ and $\sigma_W$) given
in km\,s$^{-1}$ and the scale height ($Z_0$) expressed in pc. 
Individual absolute magnitudes and more details on the luminosity calibration 
and the spatial and velocity distributions will be given in a forthcoming paper
to be published in Astronomy \& Astrophysics.

%%%%%%%%%%%%%%%%%%%%%%%%%%%%%%%%%%%%%%%%%%%%%%%%%%%%%%%%%%%%%%%%%%%%%%%%%%%%
\begin{table*}[ht]
 \caption{Dispersions of the intrinsic visual absolute magnitude, kinematics
and scale heights for the main groups}
 \label{tab:resul}
 \small
 \begin{center}
 \begin{tabular}{lrrrrrr}
            \multicolumn{1}{l}{Group} &
            \multicolumn{1}{c}{N$_{f}$} &
             \multicolumn{1}{c}{$\sigma_M$} &
              \multicolumn{1}{c}{$\sigma_U$} &
              \multicolumn{1}{c}{$\sigma_V$} &
              \multicolumn{1}{c}{$\sigma_W$} &
              \multicolumn{1}{c}{$Z_0$}\\
  \hline   
{He-rich} &  14 &1.2  $\pm$ 0.4 &  8.7 $\pm$ 4.3 & 7.6 $\pm$ 3.1 &
          5.0 $\pm$ 4.3 & 81   $\pm$ 76 \\
         
{He-weak}  &  58 & 0.6 $\pm$ 0.2 & 8.6  $\pm$ 1.2  &   8.2  $\pm$ 1.8  &
           3.7  $\pm$ 0.8  & 58    $\pm$ 9 \\
    
{HgMn} & 44   & 0.6 $\pm$ 0.4 & 9.3  $\pm$ 1.9 &10.9  $\pm$ 1.7 &
          5.4  $\pm$ 0.8 & 57    $\pm$ 10 \\
              
{Si} & 415  &  0.76 $\pm$ 0.09 &9.7 $\pm$ 0.3  &10.2 $\pm$ 0.5  &
	   5.8 $\pm$ 0.3  &69   $\pm$ 3  \\
           
{Si+} &  66& 0.75 $\pm$ 0.15 &14.6  $\pm$ 1.8  &9.0  $\pm$ 1.4  & 
           6.0  $\pm$ 0.7  &  49    $\pm$ 8 \\  
          
{SrCrEu}& 353& 0.76 $\pm$ 0.06& 19.1  $\pm$ 1.2& 9.5  $\pm$ 0.8&
          7.0  $\pm$ 0.4&  96    $\pm$ 8\\
               
{Am``normal''}& 781 & 0.51 $\pm$ 0.06 &  20.8 $\pm$ 1.4 &9.8 $\pm$ 0.5&
            7.4 $\pm$ 0.4 &122 $\pm$ 11\\
            
{Am ``mild''}& 184  & 0.86 $\pm$ 0.12& 20.4 $\pm$ 1.3 &9.5 $\pm$ 0.7 & 
             6.7 $\pm$ 0.8 &75 $\pm$ 8 \\
             
{$\lambda$\,Bootis}&39& 0.54 $\pm$ 0.12& 19.8 $\pm$  2.7&10.6 $\pm$ 1.4 &
             6.5 $\pm$ 1.1& 56 $\pm$ 11\\
  \hline
  \end{tabular}
  \end{center}
\end{table*}

%%%%%%%%%%%%%%%%%%%%%%%%%%%%%%%%%%%%%%%%%%%%%%%%%%%%%%%%%%%%%%%%%%%%%%%%%%%%
\begin{figure*}[p]
\psfig{figure=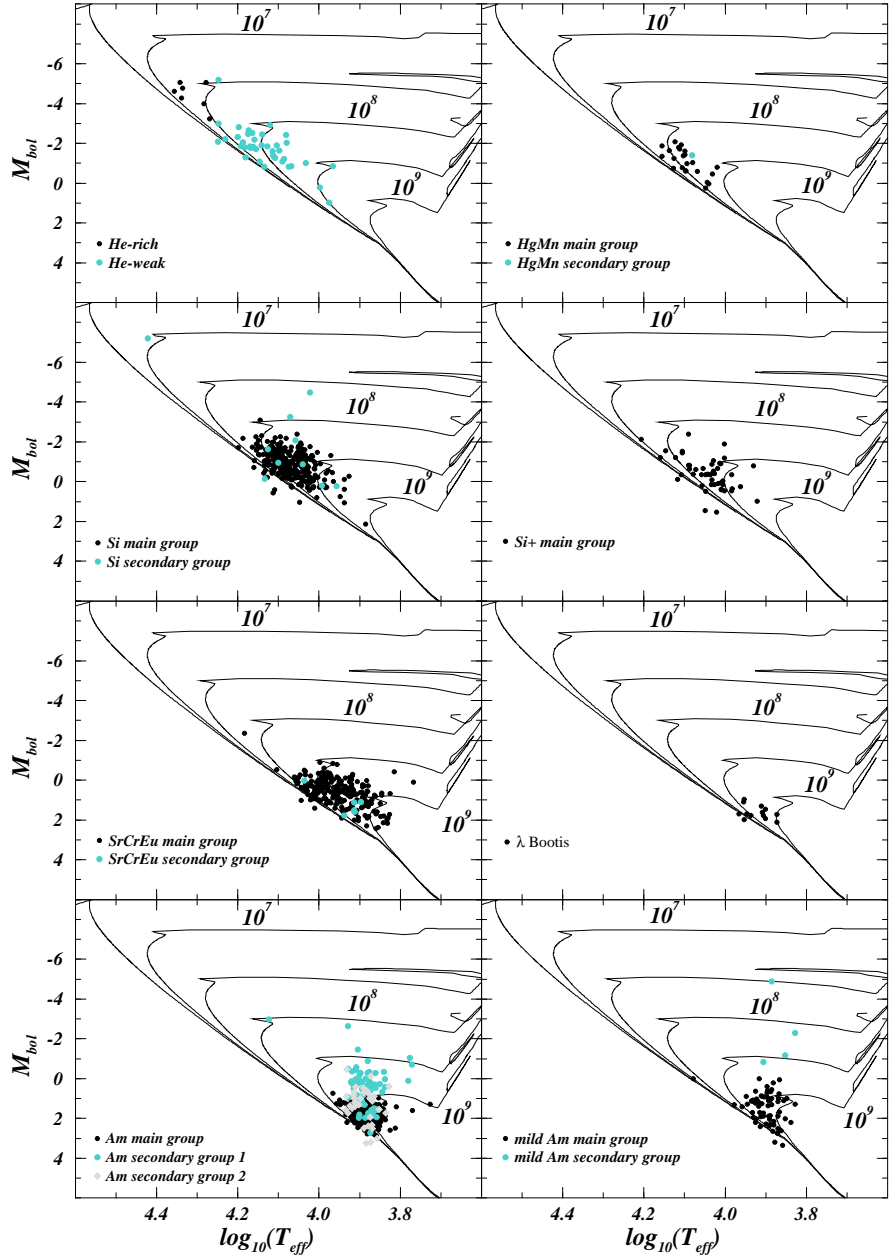,height=17cm}
 \caption{Distribution in the $[T_{\rm eff},M_{bol}]$ plane of the stars
  in the subgroups: He-rich, He-weak, HgMn, Si, Si+, SrCrEu,
  $\lambda$\,Bootis, normal Am and mild Am stars}
 \label{hr}
\end{figure*}

\vspace{-3mm}
\subsection{Absolute magnitudes}
Fig. \ref{hr} displays the position of the different subgroups in
the HR diagram ($M_{bol}$, $log\,T_{\rm eff}$).  
The bolometric absolute magnitudes $M_{bol}$ have
been calculated using the bolometric correction of St\c{e}pie\'n (1994) for 
magnetic stars, otherwise the correction of Flower (1996) has been used.
Notice that not all stars have $T_{\rm eff}$ data available. In the figure,
the isochrones of Schaller et al. (1992) for solar metallicity are also
indicated. CP stars belonging to the main groups occupy the whole width of
the main sequence. The width reaches up to 2 mag; a similar
result has been found by G\'omez et al. (1997a)  for non-peculiar stars of
 the same spectral range. The intrinsic
dispersion in absolute magnitude is rather high, varying from 0.5 to 0.8 mag 
for most of the types, except He-rich stars which spread a large range in 
luminosities.
$\lambda$\,Bootis stars are concentrated in the main sequence, but their
evolutionary status remains controversial (Pauzen, 1997). 
On kinematic grounds, their behaviour is similar to the 
one of non-peculiar stars
of the same spectral range.

Secondary groups may differ from the main groups in luminosity, in kinematics
or in both. Most of them are inhomogeneous and contain possible misclassified
and/or high-velocity objects.

\vspace{-2mm}
\subsection{Kinematics}
It is well known from the study of Bp\,-\,Ap and Am stars in associations and 
open clusters that they belong to the young disk population (see North, 1993).
Consequently, it is expected
that their spatial and velocity distributions agree with those observed for
non-peculiar main sequence 
stars of the same spectral range. For normal stars, G\'omez et al.
(1997b) obtained, using Hipparcos data, that up to 1\,Gyr $\sigma_V$ and 
$\sigma_W$ remain pratically unchanged
(10\,$\pm$\,0.5\,km\,s$^{-1}$ and 5\,-\,7\,$\pm$\,0.5\,km\,s$^{-1}$,
respectively), while $\sigma_U$ increases from 11\,$\pm$\,0.5\,km\,s$^{-1}$ at
10$^8$ years to 20\,$\pm$\,1\,km\,s$^{-1}$ at 10$^9$ years. At 2\, Gyr,
$\sigma_U$, $\sigma_V$ and $\sigma_W$ increase up to 23\,$\pm$\,1\,km\,s$^{-1}$,
14\,$\pm$\,0.5\,km\,s$^{-1}$ and 11\,$\pm$\,0.5\,km\,s$^{-1}$, respectively.
From the 
results given in Table \ref{tab:resul}, we find that, as expected, CP stars 
have the same kinematic behaviour of 
non-peculiar disk stars younger than about 1.5 Gyr. Moreover, as observed for
normal stars (Sabas, 1997; Figueras et al., 1997), the velocity field
 of CP stars shows the presence of moving groups. 
 The members of a moving group are
believed to be the result of clusters or associations in the process of 
dissociation: they are still moving with similar velocities but are
distributed over the whole sky.
 In order to identify moving groups
in the sample of CP stars, stars known to belong to associations or clusters
have been rejected as well as stars with total velocities greater than
65\,km\,s$^{-1}$. In order to compare the results for CP stars with those 
obtained by Sabas (1997) for normal stars using the same method, 
only stars brighter than V-magnitude 7.5 have been kept.
The final sample contained 467 stars.
The SEMMUL algorithm (Celeux \& Diebolt, 1986), which allows
the separation of gaussian components inside a sample (without a
priori knowledge of the number of components), has been applied using 
as input data the velocity components and their errors.
Table \ref{tab:movi} summarizes the results obtained for 
the main moving groups by Sabas (1997) from a sample of 2578 normal 
B5\,-\,F5 stars brigther than V-magnitude 7.5 and our
results for CP stars. $U_{M}$ and $V_{M}$ are the mean velocity components 
and ($\sigma_{V,M}$) the V-velocity dispersion of the moving groups
expressed in km\,s$^{-1}$, the corresponding 
mean standard errors are $\leq$\,0.5\,km\,s$^{-1}$. The number of normal (NS)
and peculiar (CP) stars found in each identified moving group is indicated
as well as the percentage of CP  stars with respect to normal stars
(given between parenthesis). Finally, the mean logarithm of the age of the 
moving groups is also given. The mean standard
error in the percentages varies between 5 and 8 \%. Notice that the 
velocity dispersion of the moving groups in the direction of the galactic 
rotation is of about 5\,km\,s$^{-1}$ (see also Figueras et al., 1997). 
%%%%%%%%%%%%%%%%%%%%%%%%%%%%%%%%%%%%%%%%%%%%%%%%%%%%%%%%%%%%%%%%%%%%%%%%%%%%
\begin{table*}[ht]
 \caption{Main moving groups found using the SEMMUL algorithm}
 \label{tab:movi}
 \small
 \begin{center}
 \begin{tabular}{lrrrcl}
            \multicolumn{1}{l}{Moving Group} &
            \multicolumn{1}{c}{$U_{M}$} &
             \multicolumn{1}{c}{$V_{M}$} &
              \multicolumn{1}{c}{$\sigma_{V,M}$} &
              \multicolumn{1}{r}{$\log(age)$} &
              \multicolumn{1}{l}{(NS\,or\,CP)}\\
  \hline
  \hline             
Pleiades& -9.7& -23.9& 4.6& 8.2&NS:\,362 \\
        &-12.7& -26.3 &3.6 & &CP:\,64 (18\%)\\
\hline        
Sirius & 10.0& 2.8 & 5.2& 8.7&NS: 388\\
       &  9.4& 3.8 & 5.9 & &CP:\,93 (24\%)\\
 \hline      
Hyades& -37.4& -15.2& 5.2 & 8.9&NS: 207 \\
      &-38.8& -19.1& 4.1&&CP:\,33 (16\%)\\
\hline	  
  \end{tabular}
  \end{center}
\end{table*}

%%%%%%%%%%%%%%%%%%%%%%%%%%%%%%%%%%%%%%%%%%%%%%%%%%%%%%%%%%%%%%%%%%%%%%%%%%%%

Secondary groups obtained with the LM method in the samples of HgMn, 
Si and SrCrEu contain a few high-velocity stars
(in total about 10). The existence of early-type
stars with apparently near solar metallicities and main sequence surface
gravities but with high-velocities and/or large distances away from the galactic
plane, constitutes a long standing anomaly in the classical picture of stellar
galactic populations (Lance, 1991). Several hypotheses have been advanced 
to explain these 
objects: ejection from the galactic plane of normal young stars,
misclassified stars like blue 
stragglers (BS) or formation as the result of accretion of gas from a merged 
satellite galaxy. Among these mechanisms, the second one is very attractive
 in the case of
CP stars because over 60\% of the BSs observed in young and intermediate 
age open clusters are found to be peculiar B\,-\,A stars (see Stryker (1993)
for details). However, it seems certain that more than one mechanism 
exists to form high-velocity early-type stars. 

Mathys et al. (1996) performed a kinematic study of
rapidly oscillating Ap stars (roAp) and found that, on kinematic grounds,
these stars are older than the non-oscillating counterparts (noAp).
We have 12 roAp and 9 noAp stars in common with their samples. Using these
stars, the velocity dispersions have been calculated. We found that both
groups have similar kinematic characteristics: $\sigma_U$, $\sigma_V$ and 
$\sigma_W$ values are 25\,$\pm$\,5\,kms$^{-1}$, 11\,$\pm$\,2\,kms$^{-1}$,
11\,$\pm$\,2\,kms$^{-1}$ and 21\,$\pm$\,5\,kms$^{-1}$,
15\,$\pm$\,4\,kms$^{-1}$, 
9\,$\pm$\,2\,kms$^{-1}$ for roAp and noAp stars, respectively. 

\vspace{-2mm}
\section{Conclusions}
\vspace{-1mm}
Our main results can be summarized as follows:\\
- Most CP stars are main sequence objects occupying the whole width of
the sequence (about 2 mag). The intrinsic dispersion
in absolute magnitude varies from 0.5 to 0.8 mag for all the groups except
He-rich stars which spread a large range in luminosities. Some Am stars in 
the secondary groups are out of the main sequence, but before 
reaching a definitive conclusion  
it will be necessary to search for possible misclassifications.\\
- From a kinematic point of view, CP stars belong to the disk population
younger than 1\,-\,1.5 Gyr. The velocity field shows the presence 
of moving groups as observed for normal stars of the same spectral range. 
In particular, the presence of Pleiades, Sirius and Hyades moving
groups has been clearly established.\\
- $\lambda$\,Bootis stars are concentrated in the main sequence. The 
definition of this type of stars is not well established (see Gerbaldi, these
proceedings). Their evolutionary status remains controversial, but
the kinematic characteristics
correspond to those of non-peculiar stars of the same spectral range.\\
- roAp and noAp stars show similar kinematic characteristics. 
\vspace{-2mm}
%%%%%%%%%%%%%%%%%%%%%%%%%%%%%%%%%%%%%%%%%%%%%%%%%%%%%%%%%%%%%%%%%%%%%%%%%%%%%%

\end{document}